\documentclass[aps,prX,superscriptaddress,reprint,longbibliography]{revtex4-1}

\usepackage{graphicx}

\renewcommand\vec[1]{{\bf #1}}

\begin{document}

\title{Dirac cones, topological edge states and non-trivial flat bands in two-dimensional semiconductors with a honeycomb nano-geometry}

\author{E. Kalesaki}
\affiliation{IEMN - Dept. ISEN, UMR CNRS 8520, Lille, France}
\affiliation{Physics and Materials Science Research Unit, University of Luxembourg, 162a avenue de la Fa\"iencerie, L-1511 Luxembourg}
\author{C. Delerue}
\email{christophe.delerue@isen.fr}
\affiliation{IEMN - Dept. ISEN, UMR CNRS 8520, Lille, France}
\author{C. Morais Smith}
\affiliation{Institute for Theoretical Physics, University of Utrecht}
\author{W. Beugeling}
\affiliation{Max-Planck-Institut f\"ur Physik komplexer Systeme, N\"othnitzer Stra\ss e 38, 01187 Dresden, Germany}
\author{G. Allan}
\affiliation{IEMN - Dept. ISEN, UMR CNRS 8520, Lille, France}
\author{D. Vanmaekelbergh}
\affiliation{Debye Institute for Nanomaterials Science, University of Utrecht}

\begin{abstract}
We study theoretically two-dimensional single-crystalline sheets of semiconductors forming a honeycomb lattice with a period below 10 nm. These systems could combine the usual semiconductor properties with Dirac bands. Using atomistic tight-binding calculations, we show that both the atomic lattice and the overall geometry influence the band structure, revealing materials with unusual electronic properties. In rock-salt Pb-chalcogenides, the expected Dirac-type features are clouded by a complex band structure. However, in the case of zinc-blende Cd-chalcogenide semiconductors, the honeycomb nano-geometry leads to rich band structures including, in the conduction band, Dirac cones at two distinct energies and non-trivial flat bands, and, in the valence band, topological edge states. These edge states are present in several electronic gaps opened in the valence band by the spin-orbit coupling and the quantum confinement in the honeycomb geometry. The lowest Dirac conduction band has $S$-orbital character and is equivalent to the $\pi-\pi^{\star}$ band of graphene but with renormalized couplings. The conduction bands higher in energy have no counterpart in graphene, they combine a Dirac cone and flat bands because of their $P$-orbital character. We show that the width of the Dirac bands varies between tens and hundreds of meV. These systems emerge as remarkable platforms for studying complex electronic phases starting from conventional semiconductors. Recent advancements in colloidal chemistry indicate that these materials can be synthesized from semiconductor nanocrystals.

\end{abstract}

\maketitle

\section{Introduction}

The interest in two-dimensional (2D) systems with a honeycomb lattice and related Dirac-type electronic bands has exceeded the prototype graphene \cite{Castro09}. Currently, 2D atomic \cite{LeLay09,Cahangirov10,Houssa11,Vogt12,Gomes12} and nanoscale \cite{Park08,Park09,Simoni10,Gibertini09,Nadvornik12} systems are extensively investigated in the search for materials with novel electronic properties that can be tailored by geometry. For example, a confining potential energy array with honeycomb geometry was created on a Cu(111) surface and it was demonstrated that the electrons of the Cu surface state have properties similar to those of graphene \cite{Gomes12}. From the same perspective, it was proposed that a honeycomb pattern with a 50-100 nm periodicity could be imposed on a 2D electron gas at the surface of a conventional semiconductor by using lithography or arrays of metallic gates \cite{Park08,Park09,Simoni10,Gibertini09,Nadvornik12}. Within the effective-mass approach, linear $E(\vec{k})$ relationships were predicted close to the Dirac points in the Brillouin zone, in analogy with graphene \cite{Park08}. The group velocity of the carriers was found to be inversely proportional to the honeycomb period and to the effective carrier mass. In order to obtain a system with sufficiently broad Dirac bands, it is thus of major importance to reduce the period of the honeycomb lattices far below 50 nm and to use semiconductors with low effective mass.

Our aim in the present work is to explore theoretically the physics of 2D semiconductors with honeycomb geometry and period below 10 nm. Electronic structure calculations using the atomistic tight-binding method \cite{Jancu98,Delerue04} attest that both the atomic lattice and the overall geometry influence the band structure. We show that the honeycomb nano-geometry not only enables the realization of artificial graphene with tunable properties but also reveals systems with non-trivial electronic structure which has no counterpart in real graphene.

We consider atomically-coherent honeycomb superlattices of rock-salt (PbSe) and zinc-blende (CdSe) semiconductors. These artificial systems combine Dirac-type electronic bands with the beneficial tunability of semiconductors under strong quantum confinement. In the case of a zinc-blende atomic lattice, separated conduction $1S$ and $1P$ Dirac cones of considerable bandwidth (10's to 100's of meV) are found, as well as dispersionless $1P$ bands. Here, $1S$ and $1P$ refer to the symmetry of the wave-functions on each node of the honeycomb. The chirality of the wave-functions with respect to a pseudo-spin is also demonstrated for both Dirac cones \cite{Castro09}. This rich electronic structure is attributed to the absence of hybridisation between $1S$ and $1P$ bands. We show that the physics for fermions in honeycomb optical lattices of cold atoms with $p$ orbitals \cite{Wu07,Sun11} could be studied in nanostructured 2D semiconductors. We point out subtle differences between the electronic structure of graphene- and silicene-type honeycomb structures. In the latter case, gaps at the Dirac points can be controllably opened and closed by an electric field applied perpendicularly to the 2D structure. In the valence band of CdSe sheets, we demonstrate the existence of topological edge states in the electronic gaps opened by the spin-orbit coupling and the quantum confinement, supporting the recent work of Sushkov {\it et al.} using envelope-function theory \cite{Sushkov13}. Our atomistic calculations even predict multiple gaps with edge states.

\section{Geometry of the 2D lattices}

\begin{figure}[!t]
\centering
\includegraphics[width=0.7\columnwidth]{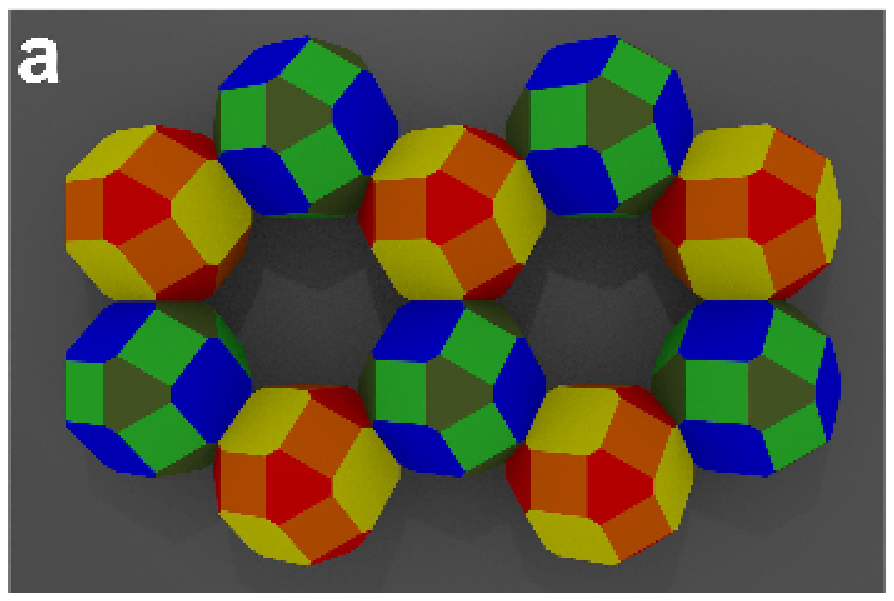}
\includegraphics[width=0.7\columnwidth]{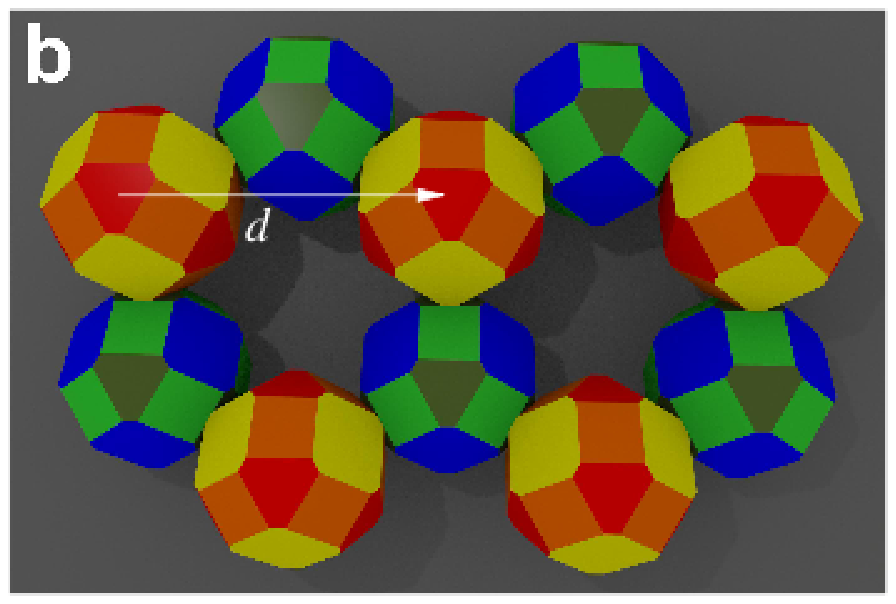}
\includegraphics[width=0.7\columnwidth]{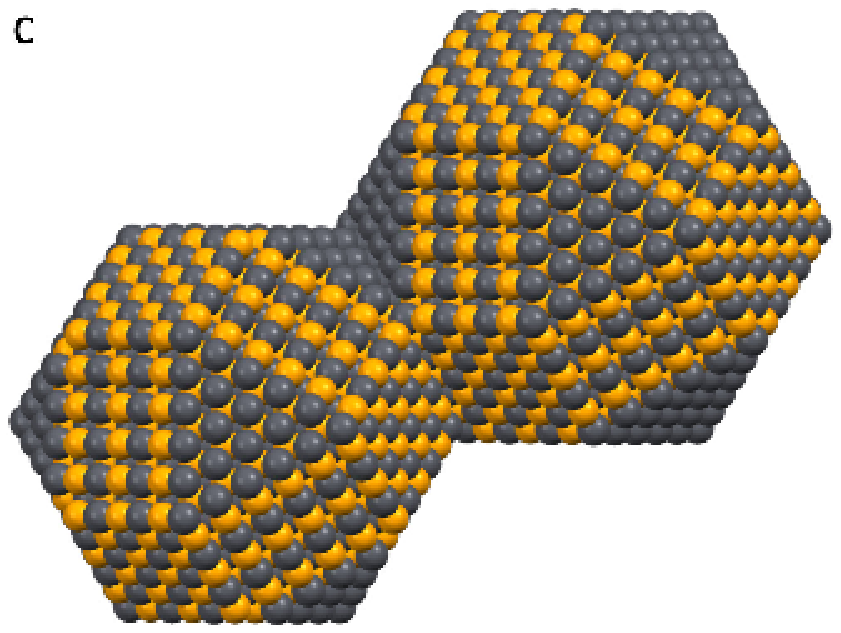}
\caption{(a,b) Block models for the self-assembled honeycomb lattices based on nanocrystals that have a truncated cubic shape. (a) Honeycomb lattice formed by atomic attachment of 3 of the 12 $\{110\}$ facets (light green/orange) corresponding to a graphene-type honeycomb structure (both sub-lattices in one plane). (b) Honeycomb lattice formed by attachment of 3 of the 8 $\{111\}$ facets (dark green/red), leading to the silicene-type configuration (each sub-lattice in a different plane). In both models, a $\langle 111 \rangle$ direction is perpendicular to the honeycomb plane as is experimentally observed. The arrow indicates the superlattice parameter $d$. (c) Top view of the unit cell of a graphene-type honeycomb lattice of PbSe nanocrystals (Pb atoms: grey; Se atoms: yellow).}
\label{fig_struc}
\end{figure}

The 2D crystals that we consider in the present work are inspired by recent experiments \cite{Evers13} discussed in Sec.~\ref{sect_exp}. These experiments consist in the synthesis in a 2D reactor plane of honeycomb sheets of PbSe by self-assembly and atomic attachment of (nearly) monodisperse PbSe colloidal nanocrystals with a truncated cubic shape \cite{Dong10,Evers13}. Due to facet-specific atomic bonding, atomically-coherent lattices with a honeycomb geometry and long-range periodicity are formed. Rock-salt PbSe lattices are transformed into zinc-blende CdSe lattices by cation exchange chemistry \cite{Son04,Evers13}. In both cases, the $\langle 111 \rangle$ axis of the atomic lattice is perpendicular to the plane of the honeycomb sheet.

The single-crystalline sheets that we consider are made of nanocrystals arranged in honeycomb structure, as shown in Fig.~\ref{fig_struc}. They can be seen as triangular lattices with a basis of two nanocrystals per unit cell forming by periodicity sub-lattices A and B. There are only two ways to assemble truncated nanocubes with a $\langle 111 \rangle$ body diagonal upright into a honeycomb lattice. The first one is to use three $\{110\}$ facets perpendicular to the superlattice plane with angles of 120 degrees, resulting in the honeycomb lattices presented in Fig.~\ref{fig_struc}a. In this structure all nanocrystal units are organized in one plane, i.e. equivalent to a graphene-type honeycomb lattice. In the second configuration, three truncated $\{111\}$ planes per nanocrystal are used for atomic contact. This results in atomically crystalline structures in which the nanocrystals of the A and B sub-lattices are centred on different parallel planes, in analogy with atomic silicene (Fig.~\ref{fig_struc}b). The separation between the two planes is $d/(2\sqrt{6}) \approx 0.2 d$ where $d$ is the superlattice parameter defined in Fig.~\ref{fig_struc}b. The unit cell of a typical graphene-type honeycomb lattice of PbSe nanocrystals is displayed in Fig.~\ref{fig_struc}c, in the case where polar $\{111\}$ facets are terminated by Pb atoms. We have also considered nanocrystals with Se-terminated $\{111\}$ facets. Further details on the lattice geometry and nanocrystal shape related to the degree of truncation are given in the Appendix~\ref{appendix_structure}.

The band structures presented below have been calculated for these quite specific geometries inspired by experiments. However, very similar results can be obtained for other honeycomb nano-geometries, as shown in Appendix~\ref{appendix_sphere} for honeycomb lattices composed of spherical nanocrystals connected by cylindrical bridges.

\section{Tight-binding methodology}

Based on the effective-mass approach, Dirac-type bands of considerable width can be expected in a nanoscale honeycomb lattice \cite{Park09}. However, a detailed understanding of the electronic structure in relation to the atomic structure and nanoscale geometry of these systems requires more advanced calculation methods. We have, therefore, calculated the energy bands of honeycomb lattices using an atomistic tight-binding method. Each atom in the lattice is described by a double set of $sp^{3}d^{5}s^{\star}$ atomic orbitals including the spin degree of freedom. We include spin-orbit interaction and we use tight-binding parameterizations (Appendix~\ref{appendix_parameters}) that give very accurate band structures for bulk PbSe and CdSe. To avoid surface states, CdSe structures are saturated by pseudo-hydrogen atoms. This is not necessary in rock-salt PbSe structures as discussed in Ref.~\cite{Allan04}. Due to the large size of the systems that we have studied (up to $6 \times 10^4$ atoms and $1.2 \times 10^6$ atomic orbitals per unit cell), only near-gap eigenstates are calculated using the numerical methods described by Niquet {\it et al.} \cite{Niquet00}.

\section{Conduction band structure of graphene-type lattices of C\lowercase{d}S\lowercase{e}}

\begin{figure}[!t]
\centering
\includegraphics[width=0.9\columnwidth]{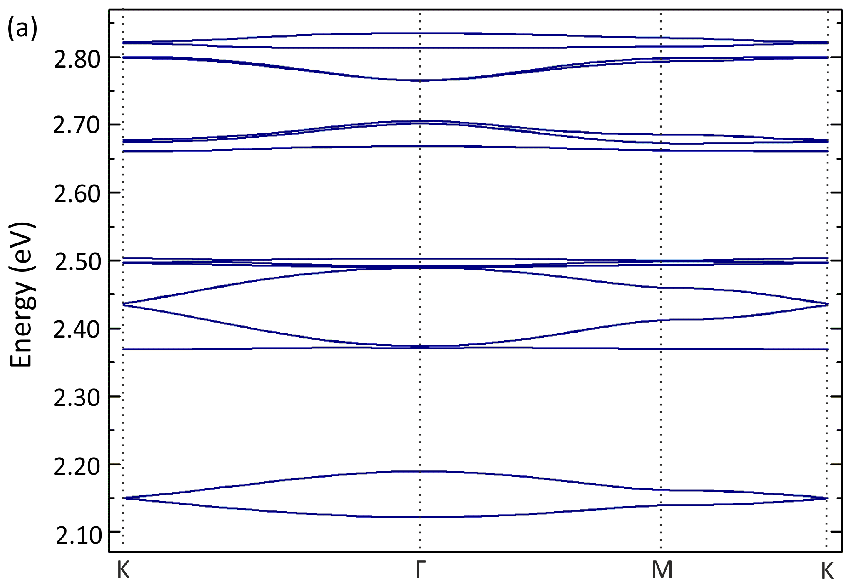}
\includegraphics[width=0.9\columnwidth]{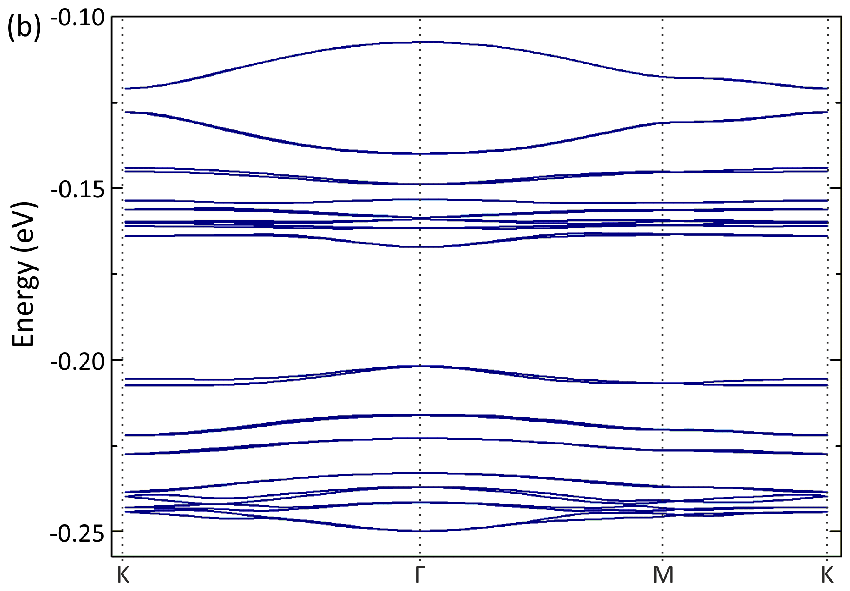}
\includegraphics[width=0.7\columnwidth]{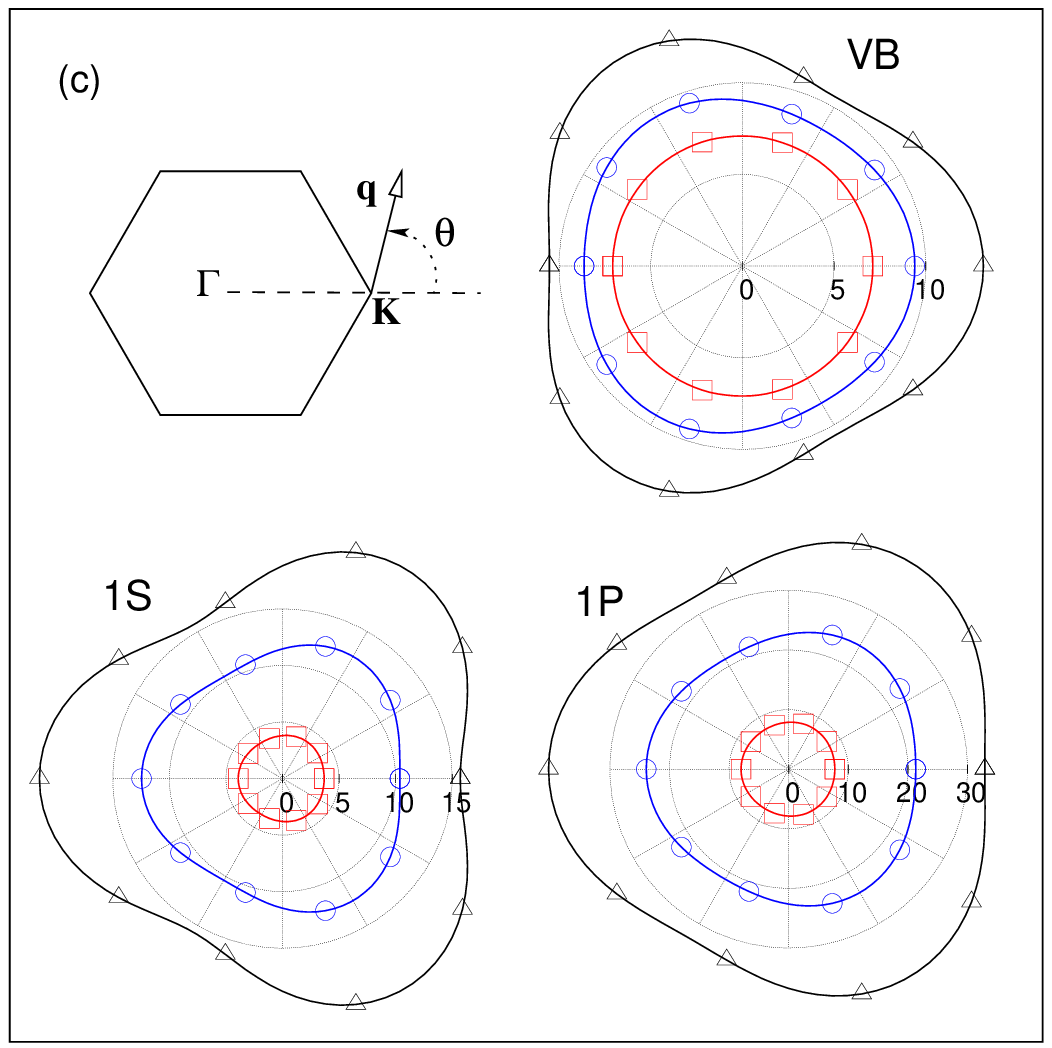}
\caption{Lowest conduction bands (a) and highest valence bands (b) of a graphene-type honeycomb lattice of truncated nanocubes of CdSe (body diagonal of 4.30 nm). The zero of energy corresponds to the top of the valence band of bulk CdSe. (c) Energy splittings (meV) between the two highest valence bands (VB), between the two 1S Dirac conduction bands (1S), and between the two 1P Dirac conduction bands (1P) calculated at $\vec{K}+\vec{q}$ and plotted using polar coordinates, i.e., energy versus the angle between $\vec{K}$ and $\vec{q}$ (squares, red curve: $|\vec{q}| = 0.05 |\vec{K}|$ ; circles, blue curve: $|\vec{q}| = 0.15 |\vec{K}|$ ; triangles, black curve: $|\vec{q}| = 0.25 |\vec{K}|$).}
\label{fig_bands_CdSe}
\end{figure}

The typical dispersion $[E(\vec{k})]$ of the highest occupied bands and the lowest unoccupied bands of a graphene-type superlattice of CdSe is shown in Fig.~\ref{fig_bands_CdSe}. The electronic structure is composed of a succession of bands and gaps due to the nanoscale periodicity. The honeycomb geometry induces periodic scattering of the electronic waves, opening gaps in particular at the center and at the edges of the superlattice Brillouin zone. 

We first discuss the simpler but extremely rich physics of the lowest conduction bands (Fig.~\ref{fig_bands_CdSe}a), which consist of two well-separated manifolds of two and six bands (four and twelve bands including spin). Strikingly, the two lowest bands have the same type of dispersion as the $\pi$ and $\pi^{\star}$ bands in real graphene; these bands are connected just at the K ($\vec{k} = \vec{K}$) and K' points of the Brillouin zone, where the dispersion is linear (Dirac points).  Moreover, in the second manifold higher in energy, four bands have a small dispersion and two others form very dispersive Dirac bands. The presence of two separated Dirac bands with Dirac points which can be experimentally accessed by the Fermi level is remarkable and has not been found in any other solid-state system. It can be understood from the electronic structure of individual CdSe nanocrystals characterized by a spin-degenerate electron state with a $1S$ envelope wave-function and by three spin-degenerate $1P$ excited states higher in energy. The two manifolds of bands arise from the inter-nanocrystal coupling between these $1S$ and $1P$ states, respectively.  Interestingly, with this nanoscale geometry, the coupling between nanocrystal wave-functions is strong enough to form dispersive bands with high velocity at the Dirac points, but small enough to avoid mixing (hybridization) between $1S$ and $1P$ states.

The systematic presence of nearly flat $1P$ bands is another remarkable consequence of the absence of $S-P$ hybridization. The existence of dispersionless bands has been predicted in honeycomb optical lattices of cold atoms with $p$ orbitals \cite{Wu07,Sun11}. In our case, two $1P$ bands are built from the $1P_z$ states perpendicular to the lattice, they are not very dispersive simply because $1P_z$-$1P_z$ ($\pi$) interactions are weak. Two other $1P$ bands ($1P_{x,y}$), respectively above and below the $1P$ Dirac band, are flat due to destructive interferences of electron hopping induced by the honeycomb geometry \cite{Wu07,Sun11} (non-trivial flat bands).

Close to the Dirac points, for $\vec{k} = \vec{K}+\vec{q}$ where $|\vec{q}| < 0.1 |\vec{K}|$, the dispersion of the 1S and 1P Dirac bands is remarkably isotropic, i.e., it does not depend on the angle between $\vec{q}$ and $\vec{K}$ (Fig.~\ref{fig_bands_CdSe}c). For larger values of $|\vec{q}|$, the Dirac cones exhibit trigonal warping due to the effect of the superlattice potential on the electrons. Trigonal deformation of the energy bands has a profound impact on the (quantum) Hall effect, interference patterns and weak localization in graphite \cite{Dresselhaus74} and bilayer graphene \cite{Kechedzhi07}. Unusual phenomena such as enhanced interference around defects and magnetically ordered exotic surfaces are also predicted at the surface of 3D topological insulators due to the hexagonal warping of the bands \cite{Fu09,Hasan09}. By analogy, similar effects could arise in the honeycomb superlattices of nanocrystals.

\begin{figure}[!t]
\centering
\includegraphics[width=0.9\columnwidth]{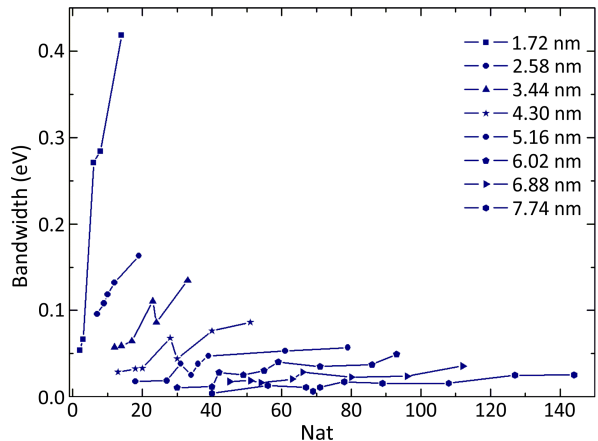}
\caption{Evolution of the bandwidth of the $1S$ Dirac band as a function of the number of atoms that form the nanocrystal/nanocrystal contact ($N_{\rm at}$). The size of the nanocrystals, which also determines the period of the honeycomb lattice, is indicated by the body diagonal of the truncated nanocube. Results are shown for the graphene geometry (bonding via the $\{110\}$ facets).}
\label{fig_width_bands_CdSe}
\end{figure}

 At $\Gamma$ ($\vec{k}=0$), the width of the $1S$ band increases not only with decreasing nanocrystal size but also with increasing number of contact atoms (Fig.~\ref{fig_width_bands_CdSe}). This can be understood by the fact that the contact area determines the electronic coupling between the nanocrystal wave-functions of adjacent sites, and acts in a similar way as the hopping parameter in atomic honeycomb lattices (see Sec.~\ref{sect_eff}). We predict a bandwidth above 100 meV for realistic configurations. Note that the width of the Dirac $1P$ bands is even considerably larger than that of the $1S$ bands. In the case of the graphene geometry, we have found that the $1S$ and $1P$ bands are always characterized by well-defined Dirac points.

Conduction bands at higher energy ($> 2.6$ eV) in Fig.~\ref{fig_bands_CdSe}a are derived from the $1D$ wave-functions of the nanocrystals. Interestingly, they also present flat bands induced by the honeycomb geometry.

\section{Non-trivial gaps in the valence band of graphene-type lattices of C\lowercase{d}S\lowercase{e}}
\label{section_topo_gap}

The twofold-degenerate conduction band of bulk CdSe near the $\Gamma$ point is mainly derived from $s$ atomic orbitals and is therefore characterized by a very weak spin-orbit coupling. It is the reason why $1S$ and $1P$ bands of CdSe superlattices are spin-degenerate and exhibit well-defined Dirac points. The situation is totally different in the valence bands of CdSe which are built from $p_{3/2}-p_{1/2}$ atomic orbitals characterized by a strong spin-orbit coupling, leading in the bulk to a splitting of 0.39 eV between heavy-hole and split-off bands at $\Gamma$.

Therefore graphene-type lattices of CdSe nanocrystals present two extremely interesting features when they are combined, 1) a honeycomb geometry, 2) a strong spin-orbit coupling, here in the valence band. In this kind of systems, the spin-orbit coupling may open a non-trivial gap and give rise to topological insulators. These systems exhibit remarkable properties at their boundaries characterized by helical edge states in the gap induced by the spin-orbit coupling \cite{Hasan10,Qi11,Beugeling12}. The edge states carry dissipation-less currents, leading in 2D systems to quantum spin Hall effect, as initially predicted for graphene by Kane and Mele  \cite{Kane05}. Whereas in graphene the spin-orbit coupling is too small to give measurable effects \cite{Castro09}, it was predicted that topological insulators with much larger gaps could be made from ordinary semiconductors on which a potential with hexagonal symmetry is superimposed \cite{Sushkov13}. We show below that honeycomb lattices of CdSe nanocrystals actually present several non-trivial gaps in their valence band.

\begin{figure}[!t]
\centering
\includegraphics[width=0.85\columnwidth]{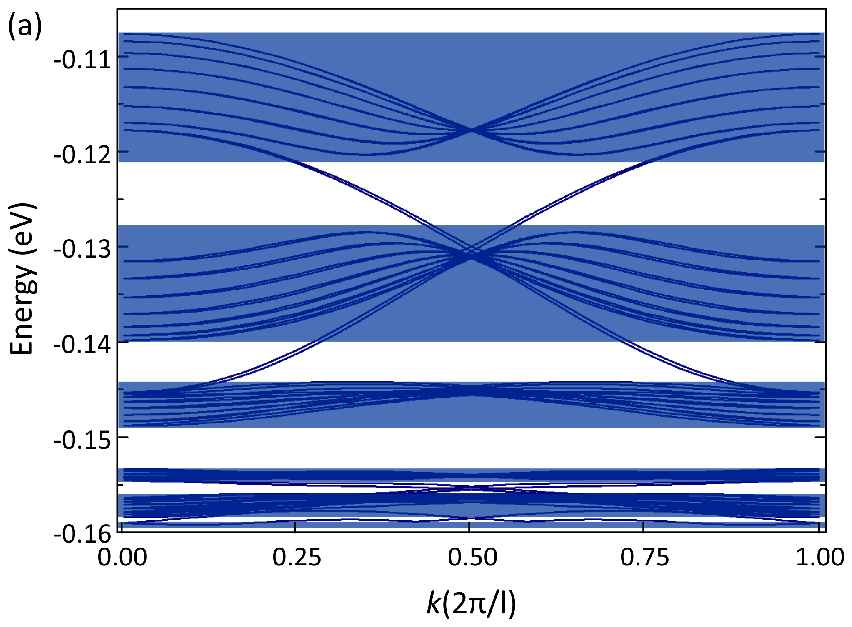}
\includegraphics[width=0.85\columnwidth]{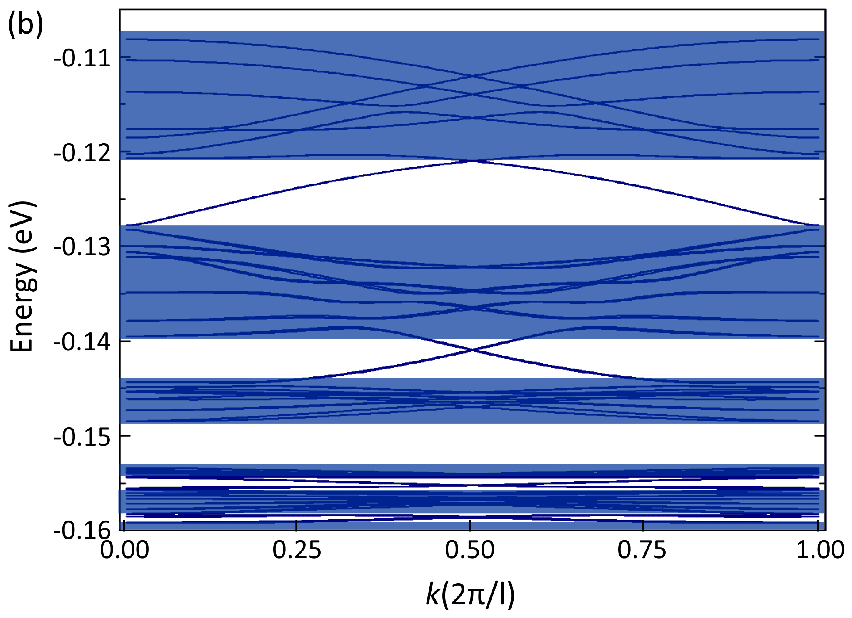}
\caption{Highest valence bands in ribbons made from a graphene-type honeycomb lattice of truncated nanocubes of CdSe (body diagonal of 4.30 nm). (a) Ribbon with zigzag edges (ribbon width = 57 nm, periodic cell length $l=8.2$ nm). (b) Ribbon with armchair edges (ribbon width = 33 nm, periodic cell length $l=14.2$ nm).  In each case, the unit cell is composed of sixteen nanocrystals (34768 atoms per unit cell). The coloured regions indicate the bands of the corresponding 2D semiconductor, i.e., those shown in Fig.~\ref{fig_bands_CdSe}b.}
\label{fig_bands_ribbon_CdSe}
\end{figure}


\begin{figure}[!t]
\centering
\includegraphics[width=0.9\columnwidth]{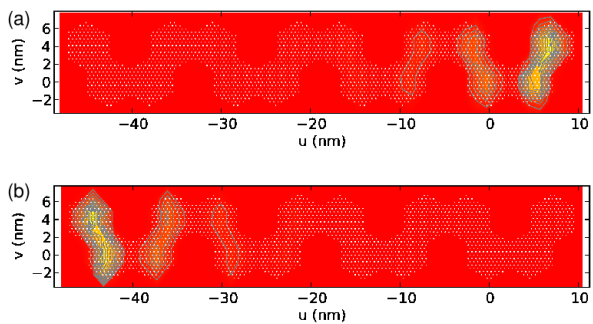}
\caption{2D plots of the wave-functions of the edge states calculated at $k = 0.3 \times 2\pi /l$ for the ribbon considered in Fig.\ref{fig_bands_ribbon_CdSe}a (energy of the states $\approx -0.123$ eV). The plots are restricted to a single unit cell of the ribbon (a: higher energy state; b: lower energy state). The vertical axis corresponds to the direction of the ribbon. More than 90\% of each wave-function is localized on the nanocrystals at the edges. The white dots indicate the atoms.}
\label{fig_wf_ribbon_CdSe}
\end{figure}

The dispersion of the valence bands in CdSe superlattices (Fig.~\ref{fig_bands_CdSe}b) is much more complex than the dispersion of conduction bands because anisotropic heavy-hole, light-hole and split-off bands are coupled by the confinement. Dirac points are not visible in spite of the honeycomb geometry. However, the highest valence bands in Fig.~\ref{fig_bands_CdSe}b roughly behave like the $\pi-\pi^{\star}$ bands in graphene, their energy-momentum relationship is isotropic close to the K point (Fig.~\ref{fig_bands_CdSe}c), but there is a large gap between the two bands. Other gaps are present at lower energy.

In the following, we investigate the topological properties of these gaps. First, we calculate the Chern numbers for the two highest valence bands of the 2D sheet \cite{Hatsugai93}. The methodology and the results are described in Appendix~\ref{appendix_chern}. The analysis of these Chern numbers demonstrates the non-trivial character of the two gaps between the highest valence bands and therefore the quantum spin Hall effect is predicted in these gaps. Second, we calculate the valence band structure of ribbons with zigzag and armchair geometry built from CdSe nanocrystals. We have considered the same honeycomb geometry as in Fig.~\ref{fig_bands_CdSe}b but for a 1D ribbon instead of a 2D sheet. The unit cell forming the ribbon by periodicity is composed of sixteen nanocrystals. Figure~\ref{fig_bands_ribbon_CdSe}a presents the 1D band structure for a ribbon with zigzag edges. We have found edge states crossing the gap between the two highest valence bands of the 2D sheet. The 2D plot of their wave-functions is shown in Fig.~\ref{fig_wf_ribbon_CdSe}. Interestingly, edge states are also present in the second gap below in energy, and others are even visible in the smaller gaps below -0.15 eV. The analysis of their wave-functions on one side of the ribbon shows that the spin is mainly oriented perpendicular to the lattice ($> 98\%$) \cite{Note_spin}, and that the direction of the spin is reversed for motion in the opposite direction ($k \to -k$). The situation is inverted at the opposite edge of the ribbon. Very similar results are obtained for ribbons with armchair edges (Fig.~\ref{fig_bands_ribbon_CdSe}b) and for ribbons in which we modify the size of the nanocrystals at the edges (not shown). The presence of helical edge states in the gaps of the 2D sheet and their robustness with respect to the edge geometry are signatures of their non-trivial topology. The multiplicity of non-trivial gaps demonstrates the variety of effects induced by the honeycomb nano-geometry on the band structure of the 2D semiconductor, contrarily to the case of a square nano-geometry \cite{Kalesaki13}.

\begin{figure}[!t]
\centering
\includegraphics[width=0.9\columnwidth]{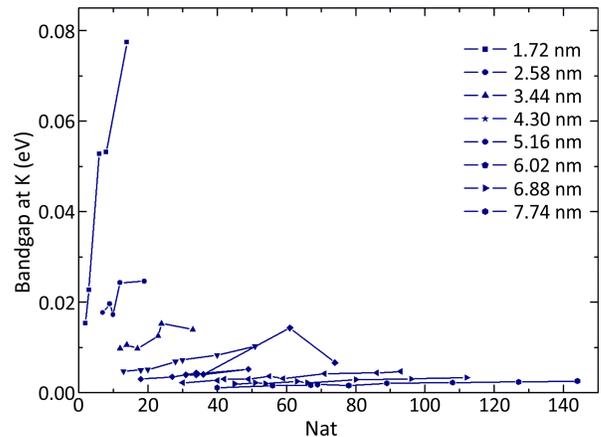}
\caption{Evolution of the energy gap between the two highest valence bands of CdSe sheets with graphene-like honeycomb geometry as a function of the number of atoms that form the nanocrystal/nanocrystal contact ($N_{\rm at}$). The size of the nanocrystals is indicated by the body diagonal of the truncated nanocube.}
\label{fig_gap_valence}
\end{figure}

The gaps between the valence bands are tunable thanks to the quantum confinement. Figure ~\ref{fig_gap_valence} shows that the gap between the two highest valence bands strongly depends on the size of the nanocrystals and increases with the number of atoms at the contact plane between neighbor nanocrystals. A gap above 10 meV is possible with nanocrystal size below 4 nm. 

\section{Pseudo-spin}

\begin{figure}[!t]
\centering
\includegraphics[width=0.9\columnwidth]{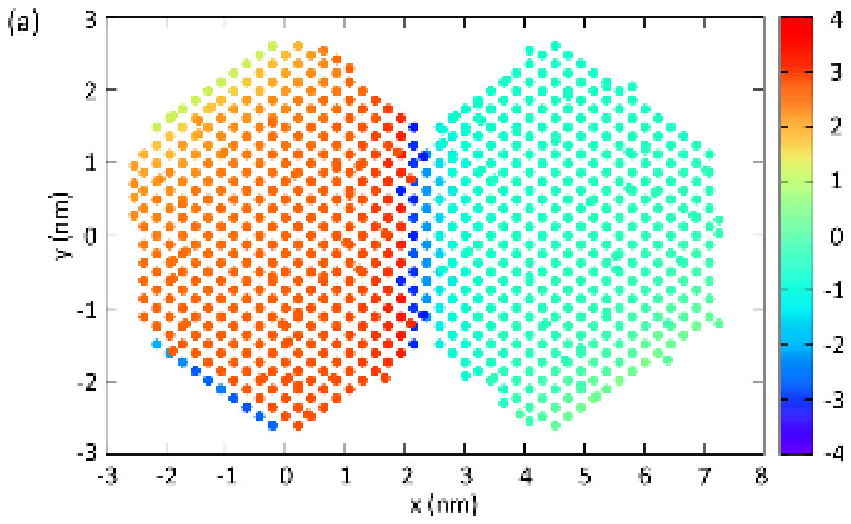}
\includegraphics[width=0.9\columnwidth]{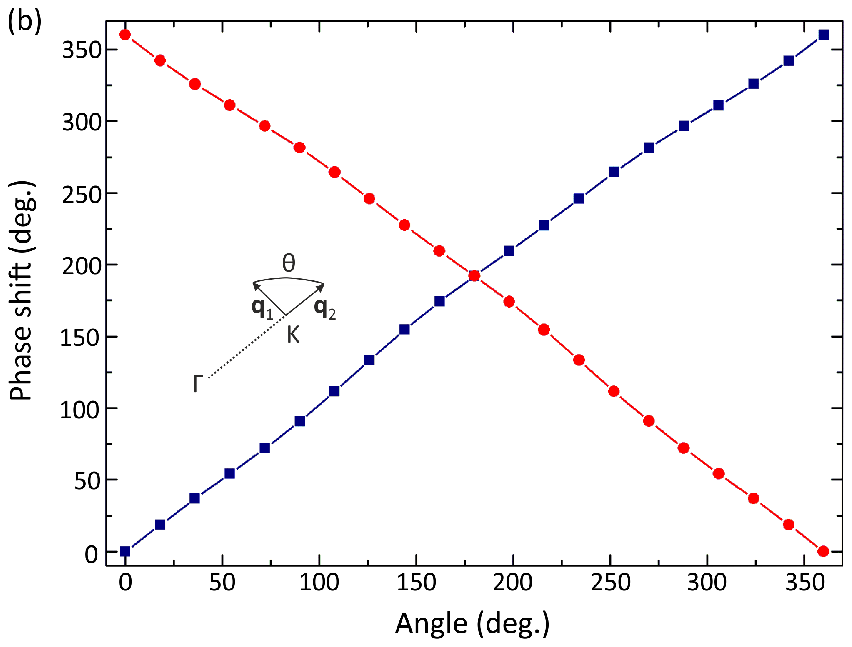}
\caption{Chirality of the $1S$ wave-functions of the CdSe honeycomb lattice compared with atomic chirality in graphene. (a) Phase shift (radians) on each atomic orbital between the electronic states of the lowest conduction band of the graphene-type honeycomb superlattice of the compound CdSe (Fig.~\ref{fig_bands_CdSe}a). The phase shift is calculated at $\vec{K}+\vec{q}_{1}$ and $\vec{K}+\vec{q}_2$ with $|\vec{q}_1|=|\vec{q}_2|=0.05|\vec{K}|$ and an angle between $\vec{q}_1$ and $\vec{q}_2$ of 3.8 radians, for each lateral atomic position in the lattice plane. (b) Difference between the phase of the wave-function at the center of nanocrystals A and B versus the angle between $\vec{q}_1$ and $\vec{q}_2$, at K (blue, square symbols) and K' (red, circular symbols). Similar results are obtained for the $1P$ Dirac cone.}
\label{fig_pseudospin}
\end{figure}

A fingerprint of the electronic states at the Dirac cones in graphene is their chirality with respect to a pseudo-spin associated with the two components of the wave-function on the two atoms of the unit cell \cite{Castro09}. Figure~\ref{fig_pseudospin} shows that the pseudo-spin is also well defined near the Dirac points in the conduction band of CdSe honeycomb lattices, in spite of the fact that each unit cell of the structure contains thousands of atoms. When we rotate the $\vec{k}$ vector around the K point, the phase shift of the wave-function is almost constant across each nanocrystal, in other words, it does not depend on the atomic orbital and its position. Note that the wave-function phase shift changes quite abruptly at the contact plane between two nanocrystals (one of sub-lattice A and one of sub-lattice B). In the lower $1S$ band, the phase difference between nanocrystals A and B is equal to the angle of rotation, and the variation is opposite at the K' point (Fig.~\ref{fig_pseudospin}b). The sign is also inverted in the upper $1S$ band. This chirality of the wave-function should reduce backscattering of the Dirac electrons, for the same reasons as in graphene \cite{Castro09}.

\section{Band structure of silicene-type lattices of C\lowercase{d}S\lowercase{e}}

\begin{figure}[!t]
\centering
\includegraphics[width=0.9\columnwidth]{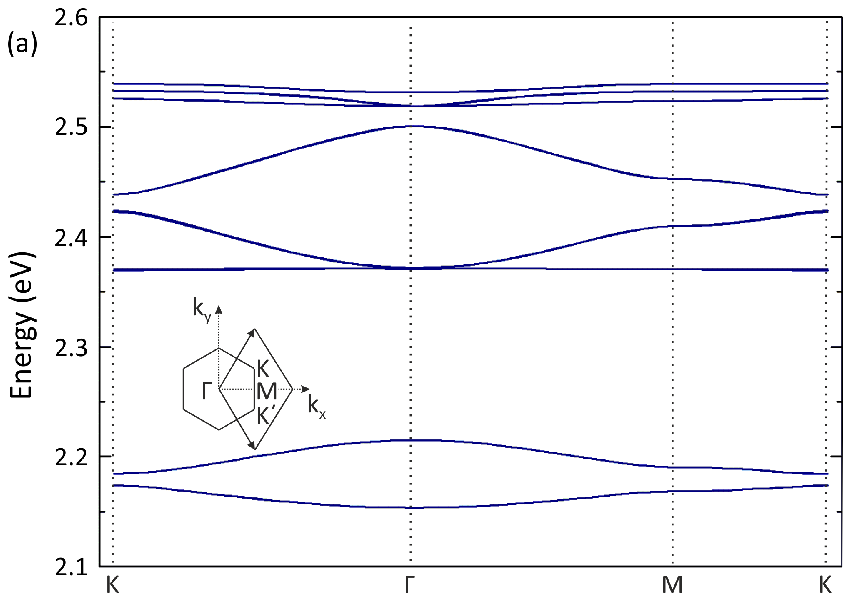}
\includegraphics[width=0.9\columnwidth]{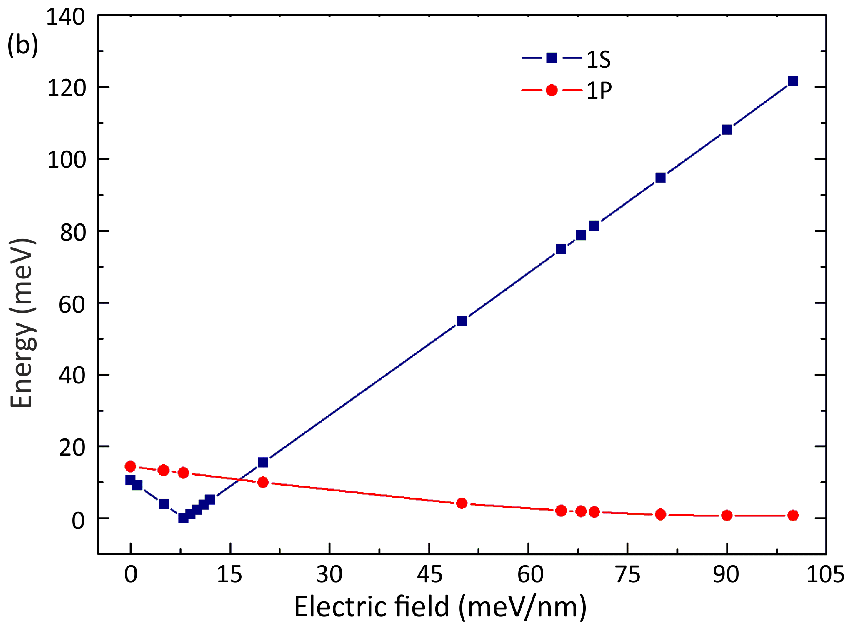}
\caption{(a) Lowest conduction bands of a silicene-type honeycomb lattice of truncated nanocubes of CdSe (body diagonal of 5.27 nm). (b) Evolution of the gap at K between the $1S$ (blue squares) and between the $1P$ (red circles) Dirac bands versus the electric field strength applied perpendicular to the honeycomb plane.}
\label{fig_bands_CdSe_sil}
\end{figure}

The band structures for honeycomb lattices of CdSe with silicene-type geometry are almost the same as for the graphene-type geometry (see Fig.~\ref{fig_bands_CdSe_sil}a for conduction bands), but there are gaps at the Dirac points due to the absence of mirror symmetry with respect to the $\{111\}$ contact plane between neighbour nanocrystals. In other words, sub-lattice A is not equivalent to sub-lattice B. The gap at K strongly depends on the size and the shape of the nanocrystals. Since in that case the nanocrystals of the A and B sub-lattices are positioned at different heights, the electronic structure is very sensitive to an electric field applied along $\langle 111 \rangle$ as shown in Fig~\ref{fig_bands_CdSe_sil}b. The gap at K between the $1S$ bands varies linearly with the field and vanishes when the potential drop between nanocrystals A and B compensates the effect of the geometrical asymmetry on the $1S$ wave-functions. The variation of the gap at K between $1P$ Dirac bands is more complex due to intermixing between $1P$ nanocrystal wave-functions, and increasing $1S-1P$ and $1P-1D$ hybridization. However, this gap tends to zero at increasing electric field.

\section{Band structure of graphene-type lattices of P\lowercase{b}S\lowercase{e}}

\begin{figure}[!t]
\centering
\includegraphics[width=0.9\columnwidth]{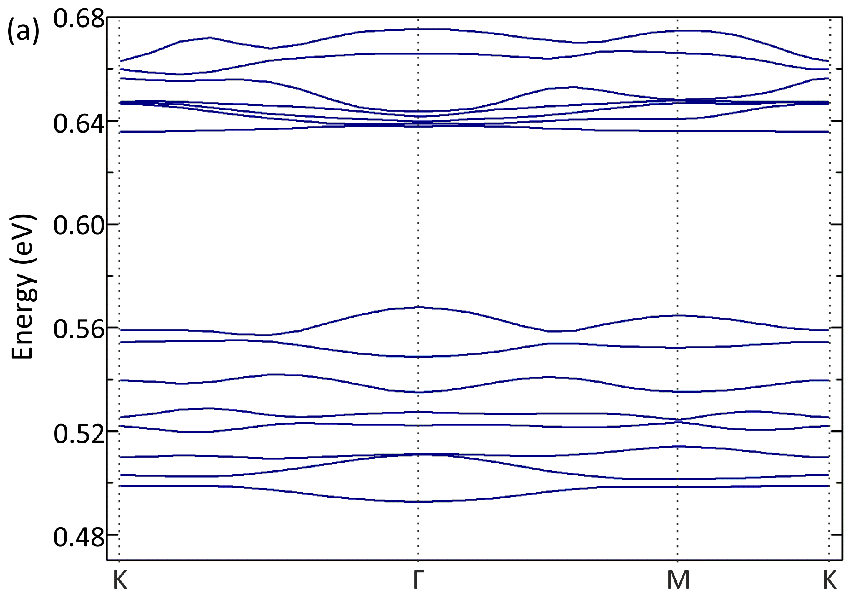}
\includegraphics[width=0.9\columnwidth]{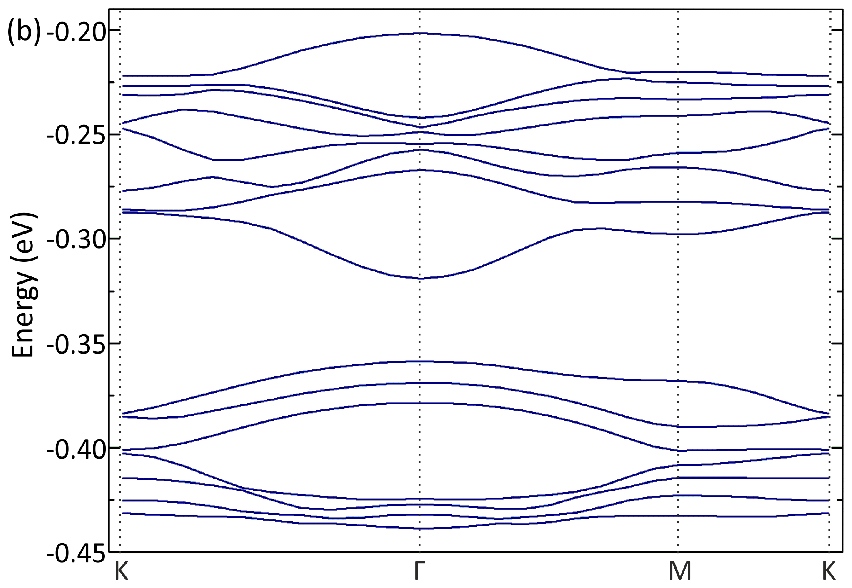}
\caption{Lowest conduction bands (a) and highest valence bands (b) of a graphene-type honeycomb lattice of truncated nanocubes of PbSe (body diagonal of 5.00 nm, $q=0.5$).}
\label{fig_bands_PbSe}
\end{figure}

Figure~\ref{fig_bands_PbSe} shows a typical band structure of a honeycomb lattice of PbSe with graphene-type geometry. The lowest conduction bands and highest valence bands are characterized by a manifold of eight bands which are formed by the $2 \times 4$ $1S$ conduction and valence wave-functions of the two PbSe nanocrystals of the unit cell. These $1S$ nanocrystal wave-functions are derived from the fourfold degenerate conduction and valence bands of bulk PbSe at the L point of the Brillouin zone (eightfold degeneracy if we include the spin). The conduction bands at higher energy and the valence bands at lower energy are derived from the 24 $1P$ nanocrystal wave-functions (only seven bands are shown for clarity). The complex dispersion of all these bands, even the $1S$ ones, shows that the coupling between the wave-functions of neighbor nanocrystals is not only governed by the symmetry of the envelope function ($1S$, $1P$) defined by the nano-geometry but also, in a subtle way, by the underlying Bloch function which depends on the originating valley. These results illustrate the fact that both the atomic lattice of the parent semiconductor and the overall nano-geometry influence the band structure.

\section{Effective tight-binding model}
\label{sect_eff}

The overall behavior of the $1S$ and $1P$ conduction bands of honeycomb lattices of CdSe nanocrystals can be interpreted using a simple (effective) tight-binding Hamiltonian of graphene or silicene in which the two "atoms" of the unit cell are described by one $s$ and three $p$ orbitals (doubled when the spin is considered). The on-site energies are $E_{s}(A)$, $E_{p_x}(A) = E_{p_y}(A)$, $E_{p_z}(A)$, and $E_{s}(B)$, $E_{p_x}(B) = E_{p_y}(B)$, $E_{p_z}(B)$ for the A and B sub-lattices, respectively. The $z$ axis is taken perpendicular to the lattice. For silicene, sub-lattice B is not in the same plane as A, the angle between the AB bond and the $z$ axis is taken to be the angle between the $[111]$ and $[11-1]$ crystallographic axes. The energy of the $p_z$ orbital is allowed to be different from the $p_x$ and $p_y$ orbitals. Following Slater and Koster \cite{Slater54}, all nearest-neighbor interactions (hopping terms) can be written in the two-center approximation as functions of four parameters ($V_{ss\sigma}$, $V_{sp\sigma}$, $V_{pp\sigma}$, $V_{pp\pi}$) plus geometrical factors. The problems for the $s$ and $p$ bands are separable when $V_{sp\sigma} = 0$. We have found that the results of full tight-binding calculations for all 2D crystals considered are only compatible with small values of $V_{sp\sigma}$, i.e., by a small hybridization between $s$ and $p$ orbitals.

\begin{figure}[!t]
\centering
\includegraphics[width=0.8\columnwidth]{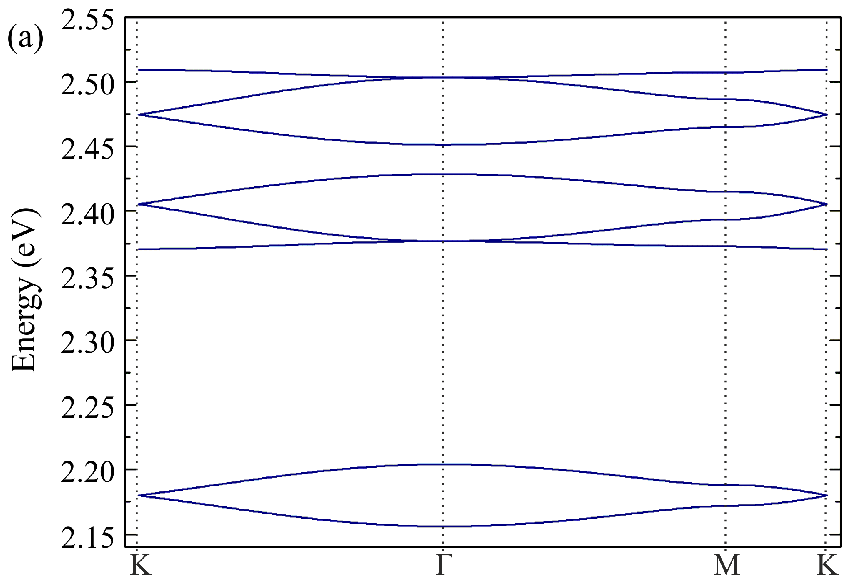}
\includegraphics[width=0.8\columnwidth]{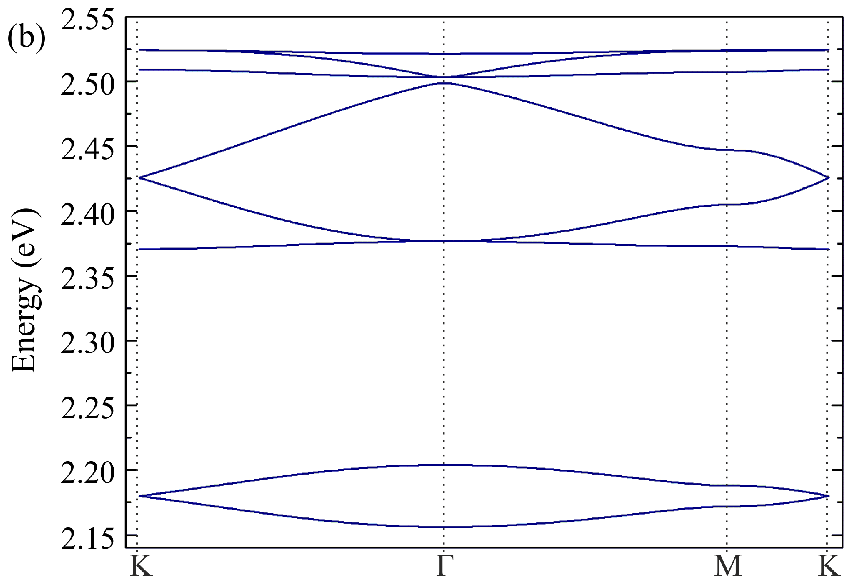}
\includegraphics[width=0.8\columnwidth]{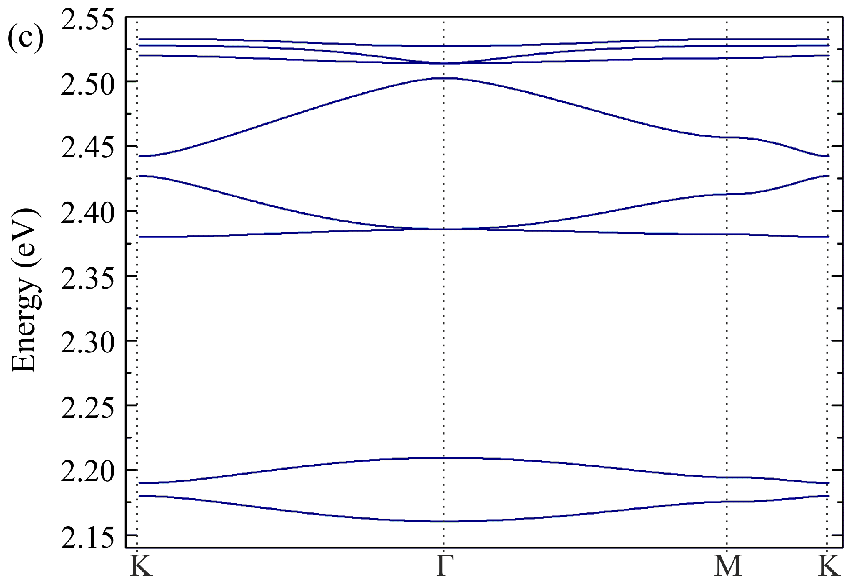}
\caption{Band structure calculated for a silicene-type lattice using the effective tight-binding Hamiltonian for three sets of parameters. (a) $E_{s}(A,B) = 2.18$ eV, all the $E_{p} = 2.44$ eV, $V_{ss\sigma} = -8$ meV, $V_{sp\sigma} = 0$ meV, $V_{pp\sigma} = 50$ meV, $V_{pp\pi} = -2$ meV. (b) Same as (a) but with $E_{p_z}(A,B) = 2.51$ eV. (c) $E_{s}(A) = 2.18$ eV, $E_{s}(B) = 2.19$ eV, $E_{p_x}(A) = E_{p_y}(A) = 2.44$ eV, $E_{p_x}(B) = E_{p_y}(B) = 2.46$ eV, $E_{p_z}(A) = 2.51$ eV, $E_{p_z}(B) = 2.52$ eV, $V_{ss\sigma} = -8$ meV, $V_{sp\sigma} = 0$ meV, $V_{pp\sigma} = 50$ meV, $V_{pp\pi} = -2$ meV.}
\label{fig_effTB}
\end{figure}

Typical band structures for the silicene geometry are shown in Fig~\ref{fig_effTB}. Figure~\ref{fig_effTB}a corresponds to a case where the on-site energies are the same on the sub-lattices A and B, and all $p$ orbitals have the same energy. Dirac points are obtained at three energies, one in the $s$ bands, two in the $p$, and no gap formed at K. This situation has not been found for more realistic band structure calculations. One Dirac $p$ band is suppressed when we slightly shift the $p_z$ energy with respect to the $p_x$, $p_y$ energies (Fig~\ref{fig_effTB}b). Finally, a small gap is opened at the two Dirac points when the orbital energies in sub-lattice A are not the same as in B (Fig~\ref{fig_effTB}c). Figure~\ref{fig_effTB}c is in close agreement with Fig.~\ref{fig_bands_CdSe_sil}a highlighting the suitability of the present effective Hamiltonian to describe the obtained results.

\section{Experimental perspective}
\label{sect_exp}

Several methods can be proposed to confine semiconductor carriers in a honeycomb geometry. The simplest method would be to use a hexagonal or honeycomb array of a metal acting as a geometric gate that would electrostatically force the carriers in a honeycomb lattice. Theoretical and experimental work with GaAs/GaAlAs has been recently reported \cite{Gibertini09,Simoni10,Sushkov13}. The advantage of this approach is the substantial theoretical and experimental knowledge already attained in the field. However, due to the relatively large period of the honeycomb lattice, the width of the bands is limited to a few meV \cite{Park09}. A second method would be to prepare suitable templates by lithography, and use these to grow semiconductor lattices with wet-chemical or gas-phase methods (chemical vapor deposition and molecular beam epitaxy). General concerns relate to the crystallinity of the 2D system prepared with a honeycomb geometry. Our tight-binding calculations unambiguously show that the electronic structure is determined both by the periodicity of the 2D semiconductor in the nanometer range and the atomic lattice.  Recently, a third approach to prepare semiconductors with a honeycomb geometry came from a rather unexpected corner. Evers {\it et al}. showed that atomically-coherent sheets of a PbSe semiconductor can be prepared by oriented attachment of colloidal nanocubes \cite{Evers13}. Moreover, the as-prepared systems can be transformed into zinc-blende CdSe by Cd-for-Pb ion exchange, preserving the nano-geometry \cite{Boneschanscher}. The systems showed astounding geometrical order and a well-defined atomic structure.  A challenge with theses systems will be to incorporate them in a field-effect transistor, such that the transport properties can be measured. As with the template-approach, the effect of defects in the atomic lattice and disorder in the superimposed honeycomb geometry, electronic doping, and the surface termination of the 2D honeycomb semiconductors are issues that need to be further addressed.

\section{Summary and future directions}

In conclusion, we have shown that atomically-coherent honeycomb lattices of semiconductors with period below 10 nm present very interesting band structures, which are defined both by the properties of the parent semiconductor and the nano-geometry. We have calculated the electronic structure of these materials using an atomistic tight-binding method, considering graphene- and silicene-type lattices obtained by the assembling of semiconductor nanocrystals. In the case of honeycomb lattices of CdSe, we predict a rich conduction band structure exhibiting non-trivial flat bands and Dirac cones at two experimentally reachable energies. These bands are derived from the coupling between nearest-neighbor nanocrystal wave-functions with $1S$ and $1P$ symmetry. The formation of distinct Dirac cones is possible because of the weak hybridization between $1S$ and $1P$ wave-functions under the effect of the strong quantum confinement. We also predict the opening of non-trivial gaps in the valence band of CdSe sheets due to the effect of the nano-geometry and the spin-orbit coupling. Several topological edge states are found in these gaps. The possibility to have multiple Dirac cones, non-trivial flat bands and topological insulating gaps in the same system is remarkable.

Recent experiments strongly suggest that the synthesis of such single-crystalline sheets of semiconductors is possible using facet-specific attachement of nanocrystals \cite{Evers13}. Therefore, numerous directions are open for the theoretical and experimental investigation of these systems. The electronic structure and carrier transport can be studied by local scanning tunnelling microscopy and spectroscopy, and in a field-effect transistor. With illumination and/or gating the conduction band can be controllably filled with electrons up to several electrons per nanocrystal \cite{Roest02,Yu03,Talapin05}, allowing for the Fermi level to cross the Dirac points. Our work provides evidence for non-trivial flat $1P$ bands, and the lowest one can be reached at a nanocrystal filling between 2 and 3 electrons. Electron-electron interactions should then play a crucial role and may lead to Wigner crystallization \cite{Wu07}. We remark that the physics of honeycomb lattices of $p$ orbitals is largely unexplored. It will be particularly attractive to consider honeycomb lattices of semiconductors with even stronger spin-orbit coupling, in which we expect the emergence of new electronic (topological) phases.

Finally, it is important to realize that all these interesting properties (Dirac cones, non-trivial flat bands and topological edge states) originate from the honeycomb geometry. For instance, the band structures of atomically-coherent square semiconductor superlattices that we discussed recently \cite{Kalesaki13}, although of interest in their own respect, do not contain any of the Dirac-based quantum electronic properties of similar semiconductors with a honeycomb nano-geometry, discussed here.

\begin{acknowledgments}
This work has been supported by funding of the French National Research Agency [ANR, (ANR-09-BLAN-0421-01)], Netherlands Organisation for Scientific Research (NWO) and FOM [“Control over Functional Nanoparticle Solids” (FNPS)]. C.D. was visiting professor at the Debye Institute for Nanomaterials Science, Utrecht University, at the time of this research.

\end{acknowledgments}

\appendix

\section{Details on the geometry of the honeycomb lattices}
\label{appendix_structure}

\subsection{Formation of the nanocrystals}

The nanocrystals are built from PbSe nanocubes on which six $\{100\}$, eight $\{111\}$ and twelve $\{110\}$ facets are created by truncation. The positions of the vertices of the nanocrystal shape are given by $P[\pm 1, \pm (1-q), \pm (1-q)]$ where $[\pm 1,\pm 1,\pm 1]$ indicate the position of the six corners of the original nanocube, $q$ is the truncation factor, and $P$ represents all the possible permutations. We have considered realistic shapes corresponding to $q$ between 0.25 and 0.5 \cite{Evers13}. The truncated $\langle 111 \rangle$-oriented nanocubes are assembled into two types of honeycomb lattices  with a graphene-like or silicene-like shape as described below.

\subsection{Geometry of graphene-type supercells}

Two nanocrystals (A and B) are attached along the $\langle 110 \rangle$ direction (perpendicular to the $\langle 111 \rangle$ axis) to define a unit cell, forming by periodicity an atomically-coherent honeycomb lattice. Each nanocrystal is defined by the same number of bi-planes of atoms in the $\langle 110 \rangle$ direction but there is an additional plane of atoms shared between neighboring nanocrystals in order to avoid the formation of wrong bonds. The length of the two vectors defining the superlattice is $(2n+1) a \sqrt(6)/2$ where $n$ is an integer and $a$ is the lattice parameter (0.612 nm for PbSe, 0.608 nm for CdSe). Since the atomistic reconstruction at the contact plane between neighboring nanocrystals is not precisely known, we have also considered structures in which we have slightly enlarged each inter-nanocrystal junction by one line of atoms on each side of the $\{110\}$ facets.

\subsection{Geometry of silicene-type supercells}

The two nanocrystals A and B are attached along a $\{111\}$ facet to define a unit cell and an atomically-coherent honeycomb lattice is formed by periodicity. Note however that the A and B nanocrystal sub-lattices are each located in a different plane. The length of the superlattice vectors (superlattice parameter) is $d = ma/ \sqrt{2}$ where $m$ is an integer. The separation between the two planes A and B is $ma/(4 \sqrt{3})$. Similar to the graphene-type lattices, we have also considered structures in which the $\{111\}$ facets at the contact plane between neighboring nanocrystals are enlarged.

\section{Tight-binding parameters}
\label{appendix_parameters}

\begin{table}  
  \caption{Tight-binding parameters (notations of Slater and Koster \cite{Slater54}) for zinc-blende CdSe in an orthogonal $sp^{3}d^{5}s^{\star}$ model. $\Delta$ is the spin-orbit coupling. (a) and (c) denote the anion (Se) and the cation (Cd), respectively.}
  \label{table_param}
  \begin{tabular}{llll}
    \hline
    \hline
    \multicolumn{4}{c}{Parameters for CdSe (eV)} \\
    \hline
    $E_s(\rm a)$ & -8.065657 & $E_s(\rm c)$ & -1.857148 \\
    $E_p(\rm a)$ & 4.870028 & $E_p(\rm c)$ & 5.613460 \\
    $E_{d_{xy}}(\rm a)$ & 15.671502 & $E_{d_{xy}}(\rm c)$ & 16.715749 \\
    $E_{d_{x^2-y^2}}(\rm a)$ & 15.232107 & $E_{d_{x^2-y^2}}(\rm c)$ & 20.151047 \\
    $E_{s^*}(\rm a)$ & 15.636238 & $E_{s^*}(\rm c)$ & 20.004452 \\
    $\Delta (\rm a)$ & 0.140000 &  $\Delta (\rm c)$ & 0.150000 \\
    $V_{ss \sigma}({\rm ac})$ & -1.639722 & $V_{s^*s^* \sigma}({\rm ac})$ & -1.805116 \\
    $V_{ss^* \sigma}({\rm ac})$ & 1.317093 & $V_{ss^* \sigma}({\rm ca})$ & 0.039842 \\
    $V_{sp \sigma}({\rm ac})$ & 3.668731 & $V_{sp \sigma}({\rm ca})$ & 1.885956 \\
    $V_{s^*p \sigma}({\rm ac})$ & 0.978722 & $V_{s^*p \sigma}({\rm ca})$ & 1.424094 \\
    $V_{sd \sigma}({\rm ac})$ & -0.890315 & $V_{sd \sigma}({\rm ca})$ & -1.007270 \\
    $V_{s^*d \sigma}({\rm ac})$ & 0.906630 & $V_{s^*d \sigma}({\rm ca})$ & 2.472941 \\
    $V_{pp \sigma}({\rm ac})$ & 4.430196 & $V_{pp \pi}({\rm ac})$ & -0.798156 \\
    $V_{pd \sigma}({\rm ac})$ & -2.645560 & $V_{pd \sigma}({\rm ca})$ & -1.296749 \\
    $V_{pd \pi}({\rm ac})$ &  0.028089 & $V_{pd \pi}({\rm ca})$ & 2.295717 \\
    $V_{dd \sigma}({\rm ac})$ & -2.480060 & $V_{dd \pi}({\rm ac})$ & 2.393224 \\
    $V_{dd\delta}$ &  -1.373199 \\

    \hline
    \multicolumn{4}{c}{Parameters for Cd-H and Se-H (eV)} \\
    $E_H$ & 0.000000 \\
    $V_{ss\sigma}$ & -35.69727 & $V_{sp\sigma}$ & 61.82948 \\
    \hline
    \hline
  \end{tabular}
\end{table}

We consider a double basis of $sp^{3}d^{5}s^{\star}$ atomic orbitals for each Pb, Cd or Se atom, including the spin degree of freedom. For PbSe, we use the tight-binding parameters as given in Ref.~\cite{Allan04}. Due to the lack of $sp^{3}d^{5}s^{\star}$ tight-binding parameters for zinc-blende CdSe, corresponding data were derived and are presented in Table~\ref{table_param}. The use of a $sp^{3}d^{5}s^{\star}$ basis allows us to get a very reliable band structure for bulk materials compared to {\it ab initio} calculations and available experimental data: see Ref.~\cite{Allan04} for PbSe. Hence, these parameters can be safely transferred to predict the electronic structure of semiconductor nanostructures \cite{Delerue04}.

The effects of the electric field on the electronic structure are calculated assuming that the field inside the superlattice is uniform. In this approximation, the calculation of the screening in such a complex system is avoided. The main conclusions of these calculations, i.e., the strong evolution of the bands with the applied field and the possibility to close or open the gap at the $1S$ Dirac point in silicene-type lattices, do not depend on this approximation.

\section{Results for another type of honeycomb nano-geometry}
\label{appendix_sphere}

\begin{figure}[!t]
\centering
\includegraphics[width=0.8\columnwidth]{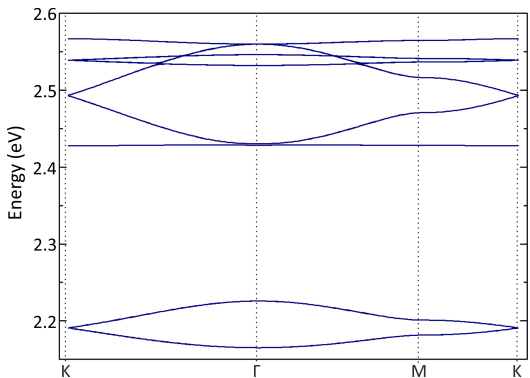}
\caption{Lowest conduction bands of a graphene-type honeycomb lattice of spherical nanocrystals of CdSe (diameter = 4.7 nm). Nearest-neighbor nanocrystals are connected by cylindrical bridges of CdSe (diameter of the cylinders = 2.3 nm).}
\label{fig_sph}
\end{figure}

It is also possible to build a honeycomb lattice using spherical nanocrystals. We set the distance between nearest-neighbor nanocrystals equal to the diameter, i.e., the spheres are tangential. Between each pair of neighbors, we add a cylinder of atoms which serves as a bond. We predict that the electronic bands for these 2D crystalline sheets are very close to those obtained with truncated nanocubes. A typical result for the conduction band is shown in Fig.~\ref{fig_sph} where the two Dirac cones and the flat bands are present. We have found similar band structures for diameters of the cylinders up to 80\% of the diameter of the spheres, demonstrating the robustness of the results. The width of the Dirac bands is related to the strength of inter-nanocrystal bonds, i.e., to the diameter of the cylinders. The non-trivial flat bands are easily identified in Fig.~\ref{fig_sph} since they are connected to the $1P$ Dirac band. Similar results are also obtained for silicene-type geometry (not shown).

\section{Calculation of the Chern numbers}
\label{appendix_chern}

We have calculated the Chern numbers of the highest valence bands following the methodology proposed by Fukui {\it et al.} \cite{Fukui05}. For each band $n$, we calculate the wave-functions $| n,\vec{k} \rangle$ on a $N \times N$ grid within the Brillouin zone, i.e., $\vec{k} = m_{1} \vec{a}_{1}^{*}/N + m_{2} \vec{a}_{2}^{*}/N$ where $\vec{a}_{1}^{*}$ and $\vec{a}_{2}^{*}$ are the reciprocal lattice vectors ($m_{1},m_{2} = 0,...,N-1$). In order to specify the gauge, we make the transformation $| n,\vec{k} \rangle \to |\langle n,\vec{k} | 0 \rangle|^{-1} \langle n,\vec{k} | 0 \rangle  | n,\vec{k} \rangle$ where $|0 \rangle$ is an arbitrary state which was chosen as constant over the unit cell. The lattice Chern number associated with the $n$th band is calculated as

\begin{equation}
\tilde{c}_{n} = \frac{1}{2 \pi i} \sum_{\vec{k}} \tilde{F}(\vec{k})
\label{lattice_sum}
\end{equation}

\noindent where $\tilde{F}(\vec{k})$ is defined as

\begin{equation}
\tilde{F}(\vec{k}) = \ln \left [ U_{1}(\vec{k}) U_{2} \left (\vec{k}+\frac{\vec{a}_{1}^{*}}{N} \right ) U_{1}^{-1} \left (\vec{k}+\frac{\vec{a}_{2}^{*}}{N} \right ) U_{2}^{-1}(\vec{k}) \right ]
\end{equation}

\noindent with $U_{i}(\vec{k}) = \langle n,\vec{k} | n,\vec{k}+\vec{a}_{i}^{*}/N \rangle / \left |\langle n,\vec{k} | n,\vec{k}+\vec{a}_{i}^{*}/N \rangle \right |$. $\tilde{F}(\vec{k})$ is defined within the principal branch of the logarithm ($-\pi < \tilde{F}(\vec{k})/i \le \pi$). It was shown that the lattice Chern number $\tilde{c}_{n}$ tends toward the usual Chern number $c_{n}$ in the limit $N \to \infty$ \cite{Fukui05}.

\begin{figure}[!t]
\centering
\includegraphics[width=0.8\columnwidth]{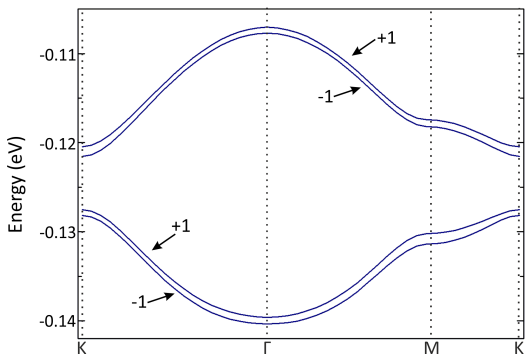}
\caption{Highest valence bands of a graphene-type honeycomb lattice of truncated nanocubes of CdSe (body diagonal of 4.30 nm) under a magnetic field $\vec{B}$ which induces Zeeman splitting of the bands ($\mu_{B} g_{S} B= 1$ meV). The corresponding bands at zero magnetic field are those shown in Fig.~\ref{fig_bands_CdSe}b. The Chern numbers of the bands are indicated on the figure.}
\label{fig_bs_zeeman}
\end{figure}

In the calculation of the Chern numbers for the superlattices, numerical and fundamental problems may arise. Numerical problems come from the size of the systems containing thousands of atoms per cell. However, the calculation of the quantities $\tilde{F}(\vec{k})$ can be done separately for each value of $\vec{k}$ on the grid, enabling powerful parallel treatment on multi-processor computers. As a consequence, the methodology proposed by Fukui {\it et al.} \cite{Fukui05} was found to be extremely efficient.

Fundamental issues are more problematic. Each band among the highest valence bands shown in Fig.~\ref{fig_bands_CdSe}b is actually composed of two almost-degenerate bands which intersect at several points of the Brillouin zone ($\Gamma$, K, M in particular). Elsewhere in the Brillouin zone, the splitting between the almost-degenerate bands is smaller than 0.2 meV in average (not visible in Fig.~\ref{fig_bands_CdSe}b). Therefore the methodology to calculate $\tilde{c}_{n}$ cannot be applied to these situations as it requires non-degenerate bands over the full Brillouin zone \cite{Fukui05}. In order to lift the degeneracies between the bands, we have studied the effect of a magnetic field applied perpendicularly to the superlattices. Since our objective is to explore the topological properties of the bands, we have only considered the Zeeman part of the coupling Hamiltonian. Figure~\ref{fig_bs_zeeman} shows that the application of a magnetic field totally splits the two highest valence bands, pushing downward (upward) the states with a majority of spin up (down) component. 

The calculated Chern numbers for the two highest manifolds of valence bands are indicated in Fig.~\ref{fig_bs_zeeman}. For each band, the sum in Eq.~(\ref{lattice_sum}) typically converges to its final value for $N \approx 30$. We have checked that $\tilde{c}_{n}$ remains constant for larger $N$ from $\approx 30$ to $100$. It also remains constant when we vary the magnetic field, from $\mu_{B} g_{S} B \approx 0.2$ meV (minimum value to remove the degenaracies) to 5 meV. It was not possible to calculate the Chern numbers for the valence bands lower in energy in this manner, because the degeneracy points remain even under an applied magnetic field. An equivalent approach to characterize the Z$_2$ topological invariant is given in Refs.~\cite{Fu07,Fukui07}.

For each valence band shown in Fig.~\ref{fig_bs_zeeman}, the sum of the Chern numbers $c_{\uparrow,n} + c_{\downarrow,n}$ for the two spin components vanishes meaning that their contribution to the Hall conductivity is zero. However, the difference $c_{\uparrow,n} - c_{\downarrow,n}$ (spin Chern number) which is linked to the spin Hall conductivity \cite{Beugeling12} does not vanish. Since the sum of the Chern numbers over all the occupied valence bands must be zero for each spin component (there is no edge state in the gap between valence and conduction bands), we deduce by subtraction that the quantum spin Hall effect must be present in the two highest gaps of the valence band, in total agreement with our previous conclusions of Sec.~\ref{section_topo_gap}.

A small remark is in place here. For the reason that the spins are not pointing exactly up or down, the difference $c_{\uparrow,n} - c_{\downarrow,n}$ is strictly speaking not well defined. However, the spins are tilted only slightly away from the vertical (deviation $< 2\%$), so that they can be treated as approximately up or down. The Chern numbers indicate that the corresponding spin ``up'' and spin ``down'' edge states counterpropagate, i.e., that the bulk gap exhibits a nonzero spin Hall conductivity while the charge Hall conductivity vanishes. The small tilt of the spins perturbs the value of the spin Hall conductivity slightly away from the quantized value, but does not destroy the topological character of the state. Thus, this state is a quantum spin Hall state in good approximation.

\bibliography{dirac}

\begin{thebibliography}{40}
\bibitem [{Not()}]{Note_spin}%
  \BibitemOpen
  \href@noop {} {}\bibinfo {note} {The spin is no longer a good quantum number
  but we define the weighted average of the spin on each atomic orbital (mean
  deviation below 5\%).}\BibitemShut {Stop}%
\end{thebibliography}

\end{document}